\documentclass{emulateapj}

\shorttitle{Molecular Gas and Galactic OB Star Formation}
\shortauthors{Luna, A. et al.}

\begin{document}

\title{Molecular Gas, Kinematics, and OB Star Formation \\
    in the Spiral Arms of the Southern Milky Way}

\author{A. Luna\altaffilmark{1,2}, L. Bronfman\altaffilmark{2},
L. Carrasco\altaffilmark{1}, and J. May\altaffilmark{2}}

\altaffiltext{1}{Instituto Nacional de Astrof\'{\i}sica Optica y
Electr\'onica, Tonantzintla, Puebla, M\'exico; aluna@inaoep.mx,
carrasco@inaoep.mx}

\altaffiltext{2}{Departmento de Astronom\'{\i}a, Universidad de
Chile, Casilla 36-D, Santiago, Chile; leo@das.uchile.cl,
jmay@das.uchile.cl}

\begin{abstract}
The rotation curve for the IV galactic quadrant, within the solar
circle, is derived from the Columbia University - U. de Chile
CO(J=1$\to$0) survey of molecular gas. A new sampling, four times
denser in longitude than in our previous analysis, is used to
compute kinematical parameters that require derivatives w/r to
galactocentric radius;  the angular velocity $\Omega(R)$, the
epicyclic frequency $\kappa(R)$,  and the parameters $A(R)$ and
$B(R)$ describing, respectively, gas shear and vorticity. The
face-on surface density of molecular gas is computed from the CO
data in galactocentric radial bins for the subcentral vicinity,
the same spectral region used to derive the rotation curve, where
the two-fold ambiguity in kinematical distances is minimum.  The
rate of massive star formation per unit area is derived, for the
same radial bins, from the luminosity of IRAS point-like sources
with FIR colors of UC H{\small II} regions detected in the
CS(J=2$\to$1) line. Massive star formation occurs preferentially
in three regions of high molecular gas density, coincident with
lines of sight tangent to spiral arms. The molecular gas motion in
these arms resembles that of a solid body, characterized by
constant angular velocity and by low shear and vorticity. The
formation of massive stars in the arms follows the Schmidt law,
$\Sigma_{MSFR} \propto [\Sigma_{gas}]^n$, with an index of $n =
1.2 \pm 0.2$.   Our results suggest that the large scale kinematics,
through shear, regulate global star formation in the Galactic disk.

\end{abstract}


\section{INTRODUCTION}

The rotation curve, describing the circular speed of rotating
material as a function of galactocentric radius, is a fundamental
tool for the study of the kinematics of our Galaxy. It is
best  derived, because of interstellar extinction, from
observations of atomic and molecular gas in radio and mm
wavelengths. The derivation involves determining the {\it terminal
velocity}, or maximum absolute radial velocity relative to
the Sun, toward lines of sight that sample the Galaxy within the
solar circle (quadrants I and IV). Such terminal velocities
correspond, assuming pure circular motion, to the tangent points
to circumferences around the galactic center, named  {\it
subcentral points}. These points subtend a circumference
that connects the solar position with the galactic center. A
detailed analysis of the rotation curve can reveal important
physical characteristics of the rotating material, such as  the
amount of shear and vorticity at each galactocentric radius. These
physical quantities regulate the gravitational stability of a
differentially rotating gaseous disk and, consequently, the large
scale distribution and properties of star formation in the
galactic disk.

The first derivation of the rotation curve for the IV galactic
quadrant that made use of the CO(J=1$\to$0) line - the best tracer
of molecular hydrogen in the interstellar medium - was presented
by Alvarez, May, \& Bronfman (1990). The spectral data used to
determine the terminal velocities were taken from the Columbia -
U. de Chile surveys \citep{ grabelsky87, bronf89}, which have a
sampling interval of 0$^{\circ}$.125 (roughly the beam size).
However, the terminal velocities in \citet{alvarez90} were
measured only every 0$^{\circ}$.5 in galactic longitude, due to
difficulties involved in the visual examination of a very large
number of spectra. A new derivation of the rotation curve, that
uses a computer search code to examine all the available spectra (
$\approx$15000), is presented here. The disk kinematic
characteristics in the IV galactic quadrant are analyzed in
detail, from this new rotation curve. These
characteristics, as a function of galactocentric radius, are
compared with the molecular gas density and with the local rate of
massive star formation.

 A proper derivation of the spiral pattern of our Galaxy
 requires knowledge of the distances to the adopted tracers.
These distances are also required to compute the masses and
luminosities of such tracers. For the gas, kinematical distances
can be obtained from radial velocity data of radio line
observations, adopting a rotation curve, under the assumption
of pure circular motions. For clouds within the solar circle, however,
there is a two-fold ambiguity in the kinematic distance, that is difficult to
circumvent and has to be resolved in a case-by-case basis. But in
the vicinity of the subcentral points such ambiguity is minimal,
since at the subcentral points themselves the kinematic distances
are univocally defined.

It is worth noting that large scale streaming motions in spiral arms,
with amplitude of $\sim$10 km/s, which produce deviations from pure
rotation, have been observed in a number of regions of the Galaxy
(Burton et al. 1988).  Streaming motions of such amplitude may
introduce uncertainties of up to 5\% in the estimation of galactocentric
radii when the streaming is along the line of sight. In such unfavorable
case, the corresponding  uncertainties in the estimated distances, for the
section of the Galaxy analyzed here, may go from of 0.6 kpc to 1.7 kpc.
In any case, for objects beyond $\sim$\,3\,kpc from the Sun, because of
optical extinction, kinematical distances are usually the only ones available.

Massive stars are formed within aggregates of molecular gas and
dust of 10$^5$-10$^6$ solar masses, about 50-100 pc in size,
which are commonly known as giant molecular clouds, or GMCs for
short. The association between OB stars and the interstellar medium
has been established through optical, infrared, and  CO observations of
GMCs close enough to be largely unaffected by extinction
(Orion, Carina, etc).  The physical conditions in GMCs control
their rates of OB star formation, and are one of the main agents that
regulate the evolution of the galactic disk \citep{evans99}.

There is a close relationship between the galactic spiral structure and
the formation of GMCs and, hence, with the formation rate of OB stars
\citep{dame86, solomon86}.  Therefore, the GMCs and the regions of
OB star formation provide a very good tool to trace the spiral arm pattern of
a galaxy.  An early description of the Milky Way spiral arm pattern was given
by \citet{gg76}, who observed the H109$\alpha$ line emitted in
H{\small II} regions associated with young massive stars.  A four arm spiral
pattern for the southern Milky Way was later proposed by \citet{cyh87}, using
a larger observational database of hydrogen recombination lines (H109$\alpha$ \&
H110$\alpha$). The four arm spiral pattern is in general agreement with that
obtained from  H{\small I} and CO large scale observations of the Galaxy
\citep{rob83, grabelsky87, bronf88, alvarez90, valle02}.

Star formation is likely to occur  in regions where the gas in the Galactic
disk is unstable to the growth of gravitational perturbations.  In a classical
paper, \citet{schm59} introduced the parametrization of the volume
density of star formation and the volume density of gas, relating
them through a power law; such parametrization, known as
"Schmidt Law'', has been studied observationally \citep{kennic89,
wong02} and explained on theoretical grounds \citep{toomre64,
tan00}. A study of the gas stability in the galactic disk must include
(a) comparison of the gas density with a critical value above which
the gaseous aggregates undergo gravitational collapse
\citep{toomre64, kennic89} and (b) examination of the gas shear rate,
that governs the process of destruction of  molecular clouds
(e.g. Kenney, Carlstrom, \& Young 1993; Wong \& Blitz 2002), presumably
through the injection of turbulent motions  \citep{maclow04}.

The link between massive star formation and kinematical conditions
in disks has been studied mostly for external spiral galaxies
\citep{aalto99, wong02, bossier03}, where the spatial resolution
that can be achieved by the observations is not as good as for the
Milky Way. The main goal of the present paper is, therefore, to
accurately describe the spiral arm structure in the {\it
subcentral vicinity} of our Galaxy, focusing on the molecular gas
kinematics, density, and on the rate of massive star formation,
with the hope of contributing to the understanding of the
formation and evolution of disk galaxies in general. The analysis
is carried out  for the IV galactic quadrant, where the spiral
structure is more evident \citep{bronf88} than in the I quadrant.
Preliminary work has been presented by \citet{cys83} and, more
recently by \citet{aluna01}.

Section (\S 2) describes the observational datasets used, the most
complete available in their kind. These data are used in section
(\S 3) to derive the rotation curve and analyze the relation
between molecular gas kinematics, molecular gas surface density,
and massive star formation rate.  The validity of Schmidt Law for
the Milky Way is analyzed in section (\S 4), and a summary of the
results is given in section \S 5.

\section{OBSERVATIONS}

The data used to derive the rotation curve and the molecular gas
surface density are part of the Columbia-U. Chile
$^{12}$CO(J=1$\to$0) surveys. These surveys provide us with the
most extensive and homogeneous observational dataset of CO
emission in the galactic disk \citep{grabelsky87, bronf89,
dame01}. The beam-size of the antenna in the CO line is
8$\arcmin$.8, and an angular sampling of 0$^\circ$.125 was
adopted. The surveys cover the entire IV galactic quadrant in
longitude, and  $\pm 2^\circ$ in latitude about the galactic
equator. The velocity resolution is 1.3\,km\,s$^{-1}$, and the
typical rms noise antenna temperature of the observations is
0.1\,K.  Main beam temperatures, $T_{MB}$, are used throughout the
analysis, obtained by dividing the antenna temperature $T^*_A$ by
the main beam efficiency, $T_{MB}=T^*_A$/$\eta_{MB}$, with
$\eta_{MB}=0.82$ \citep{bronf89}. Hereinafter we refer to the
$^{12}$CO(J=1$\to$0) line as CO. For a detailed description of the
observations see \citet{grabelsky87} and \citet{bronf89}.

The rate of massive star formation is estimated  from the integrated
FIR luminosity \citep{Boul88} of IRAS point-like sources, with FIR
colors of UC H{\small II} regions \citep{wyc89}, detected in a
CS(J=2$\to$1) survey of 1427 sources in the whole Galaxy by
Bronfman, Nyman, \& May (1996).  The CS(J=2$\to$1) emission
line requires high molecular gas densities, of $10^4$ - $10^5$\,cm$^{-3}$,
to become excited. Therefore, it  constitutes a good tracer of massive
star forming regions. The survey used here is the most complete
currently available, listing 843 massive star forming regions in
the galactic disk; the observed velocity profiles provide a good
estimator of kinematical distances to the sources.

The observations of the CS(J=2$\to$1) line toward the IV galactic
quadrant were obtained with the SEST (Swedish-ESO Sub-millimeter
Telescope) at La Silla  Observatory, in Chile, with a beam-size of
50''. Typical rms noise in the spectra is 0.1 K, at a velocity
resolution of 0.52\,kms$^{-1}$.  The radial velocity coverage is
260\,kms$^{-1}$, large enough to detect all the sources in the
radial velocity range allowed within the solar circle. Hereinafter
we refer to the CS(J=2$\to$1) transition  as CS , and  to the IRAS
point-like sources detected in CS as IRAS/CS sources. A number of
66 sources from the \citet{bronfnyman} survey are used in the
present analysis. This number was complemented with 13 sources,
undetected in the original CS survey, yet detected in a new
CS survey  with three times better sensitivity, completed with the
SEST telescope, to be published elsewhere.

The present analysis adopts the IAU recommended values for the
galactocentric distance of the Sun and for the solar circular
velocity, i.e. $R_0=8.5$\,kpc and $V_0=220$\,kms$^{-1}$
respectively \citep{kerrlynde86}. The analysis excludes galactic
longitudes from 350$^\circ$ to 360$^\circ$, because the method
used to derive the kinematical distance is highly uncertain
for that region. Furthermore, the kinematics of the central
region of the Milky Way are more complex than those of the
galactic disk \citep{sawada01}.

\section{ANALYSIS}

\subsection{Derivation of the rotation curve}

The rotation curve is derived from {\it terminal velocities} of CO
spectra at each sampled galactic longitude $l$. Since the emission
in the IV quadrant  is blue-shifted, the most negative velocity is recorded,
as well as the galactic latitude $b$ where it occurs (Fig. 1).
To define the terminal velocity, following  work by \cite{sinha78} for
H{\small I} and by \cite{alvarez90} for CO, the half intensity point of
the blue-shifted side of the emission peak with the most negative
velocity is selected (Fig. 2).  To be considered, an emission peak
is required to be larger than 4 times the noise temperature $T_{rms}$.
The emission lines thus selected have an average peak temperature
of $1.8 \pm 0.7$\,K, and an average HWHM of $3 \pm 0.8$\,kms$^{-1}$.

\begin{figure}
\epsscale{1.2} \plotone{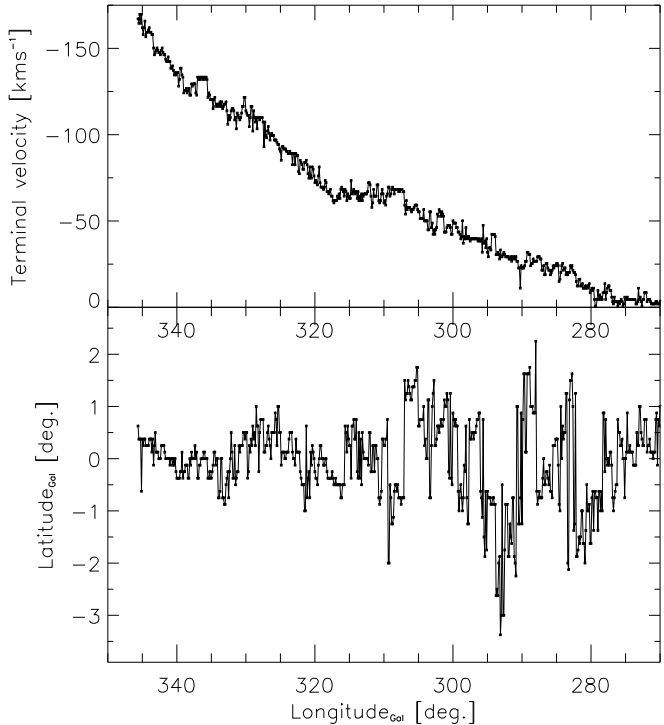} \caption{ Terminal velocities in
the IV galactic quadrant. {\it Top}: Measurements obtained every
($0.^{\circ}125$) in Galactic longitude.  {\it Bottom}: galactic
latitude where each measurement was obtained. \label{fig1}}
\end{figure}

\begin{figure}
\epsscale{1.3} \plotone{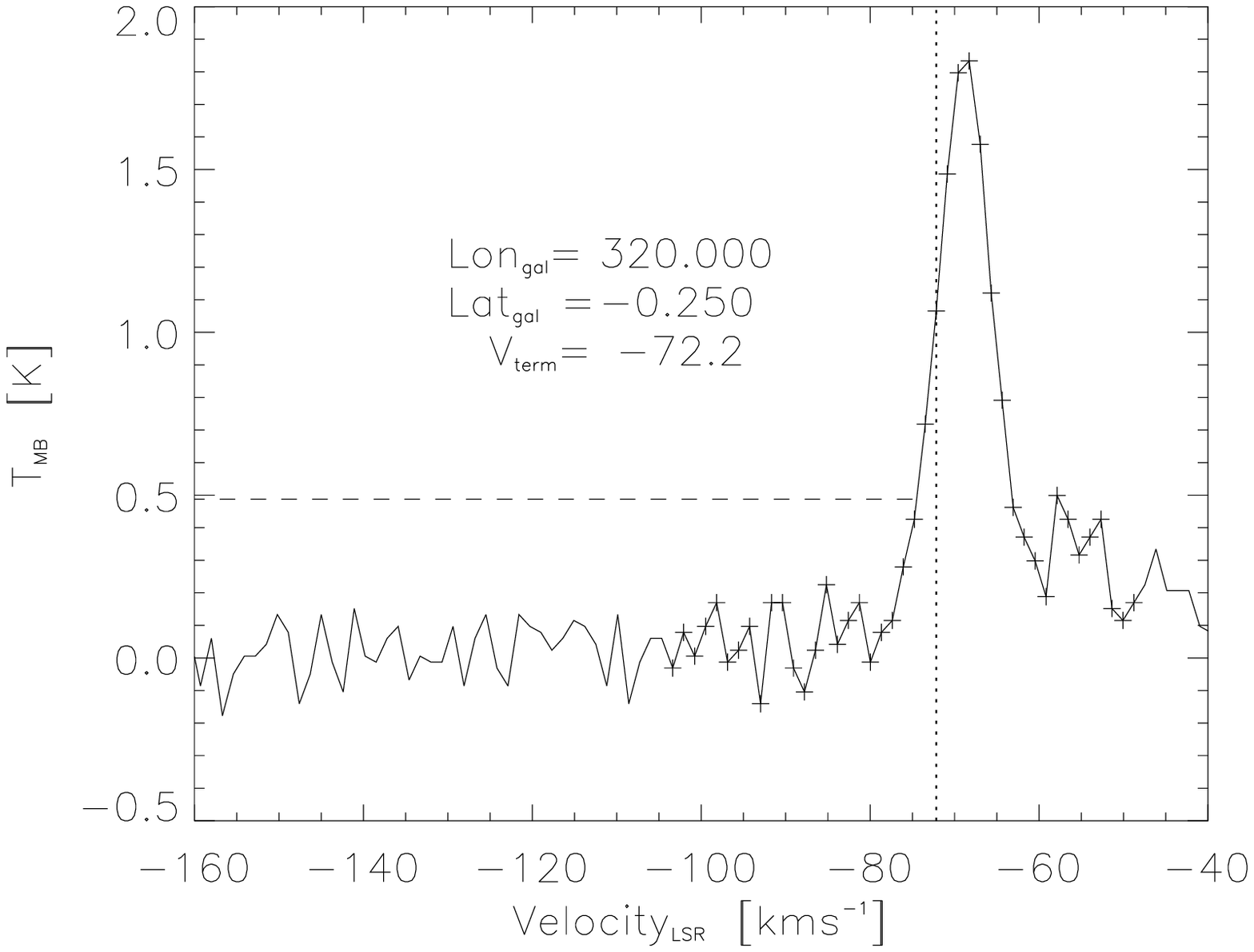} \caption{The detail of a typical
spectrum ($l=320.000$, $b=-0.250$) showing our selection criterion
in velocity and temperature. The portion of the spectral velocity
zone analyzed is marked with crosses at the measured temperature.
For this feature, the maximum temperature is $T=1.83 K$ at $v=67.8
km\,s^{-1}$. The terminal velocity value $V_{term}=-72.2
km\,s^{-1}$ is selected at the half-maximum temperature in the
blue-shifted side of the line,  and is indicated by the vertical
dotted line. A 4$\sigma$ noise level is indicated by the
horizontal dashed line. \label{fig2}}
\end{figure}

The rotation curve is obtained from the set of terminal velocities
by assigning to every longitude a subcentral point of
galactocentric distance $R = R_0 \mid \sin (l) \mid$ and
heliocentric distance $D = R_0 \cos (l)$. The rotational velocity
is then derived for each subcentral point, and assigned to the
galactocentric radius $R = R_0 \mid \sin (l) \mid$, under the
assumption of (a) pure circular motions and (b) differential
rotation with angular velocity not growing with galactocentric
radius \citep{sofuerubin01}. The rotation curve obtained, shown in
Figure 3 ({\it top}), is similar to that derived by
\citet{alvarez90}, but more detailed; the agreement for the
longitude range covered in both studies is better than
$\pm2$\,km\,s$^{-1}$ (rms). A slight difference is apparent in the
radial range $R/R_0=$ [0.56,0.62], where their analysis yields a
systematic shift in velocity of $2$\,km\,s$^{-1}$.

The distance $Z \equiv R \tan (b)$ of the subcentral point to the
galactic plane (Fig. 3 {\it bottom}) is also determined, for each
longitude, providing a simple tool to examine the $Z$ distribution
of the CO emission as a function of galactocentric radius. The
results obtained compare well with those by \citet{alvarez90},
which adopted the same method, and with those obtained from
an axisymmetric analysis of the CO emission in the IV  quadrant
\citep{bronf88}.

\begin{figure}
\epsscale{1.2} \plotone{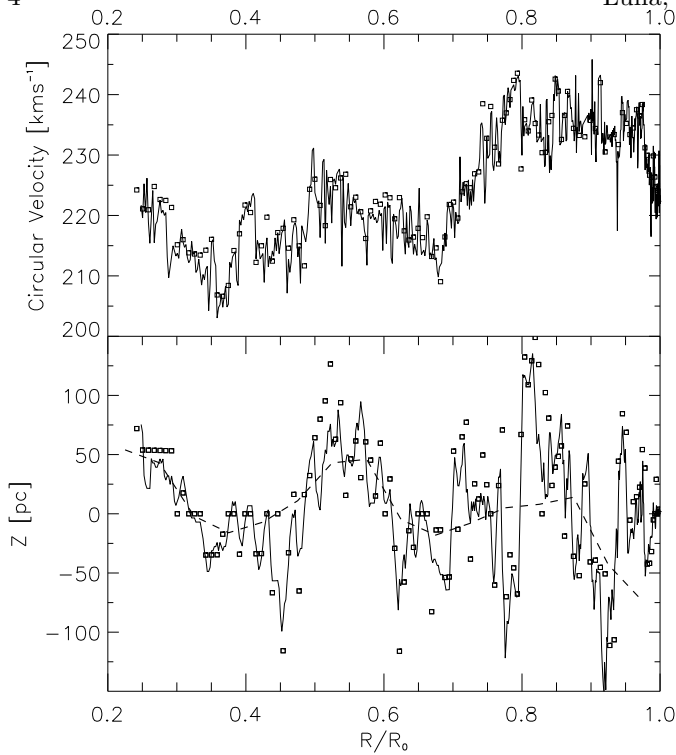} \caption{Rotation curve for the
galactic quadrant IV. {\it Top:} In our present analysis, shown by
the thin line, the longitude resolution is four times better than
in \citet{alvarez90}, shown  by small squares. The rms difference
for positions analyzed in both studies is
 2 km\,s$^{-1}$.  {\it Bottom:} Our results for the position $Z$ of the
 subcentral point  $w/r$ to the galactic equator (thin line), are
 fairly consistent  with those obtained by \citet{alvarez90}, shown
 by small squares, and with the emission centroid $Z_0$ obtained by \citet{bronf88}
 through an axisymmetric analysis of the full quadrant IV CO dataset.
(dashed line).\label{fig3}}
\end{figure}

\subsection{Disk kinematics}

The rotation curve contains vast amount of information about
the kinematics of the galactic disk \citep{byt87}. Among the
principal parameters characterizing the kinematics, derived here,
are $A(R)$ and $B(R)$, which describe the radial trends of shear
and vorticity, respectively; these parameters, when evaluated at
$R = R_0$, correspond to the well known Oort's constants.

As observed from a non-inertial coordinate frame, e.g. that
centered on the Sun (Local Standard of Rest, LSR), the balance
between Coriolis, centrifugal, and gravitational forces induce
non-circular orbits about the galactic center. These motions can
be described as small periodic oscillations superimposed onto
circular orbits. In a coordinate frame with its origin in the galactic
center and corotating with the solar orbit, these periodic
oscillations trace small ellipses called {\it epicycles}, with
an associated {\it epicyclic frequency}, $\kappa$, which roughly
accounts for deviations from circular motion. The value of $\kappa(R)$
depends, at each galactocentric radius, on the angular velocity  $\Omega(R)$:

\begin{equation}
[\kappa(R)]^2=4[\Omega(R)]^2\left[1+\frac{1}{2}\frac{R}{[\Omega(R)]}\left(\frac{d[\Omega(R)]}{dR}\right)_R\right],
\end{equation}

\noindent or, using the Oort's $A$ and $B$ parameters in their general
definition \citep{byt87},

\begin{equation}
A(R)=-\frac{1}{2}R\frac{[d\Omega(R)]}{dR} \hspace{0.3cm} ,
\hspace{0.3cm} B(R)=-\Omega(R)+A(R),
\end{equation}

\noindent it is possible to write the epicyclic frequency as:

\begin{equation}
\kappa(R)=\sqrt{-4[B(R)][A(R)-B(R)]}.
\end{equation}

A smoothed version of the rotation curve (Fig. 4 {\it top}) is
used in the calculation, in order to provide continuity for the
derivative of $\Omega(R)$ with respect to $R$ .  The size of the
smoothing box is 0.5\,kpc,  smaller than the typical width of
spiral arms. There are three galactocentric radial regions, marked
in Fig. 4 {\it top}, where $V(R)$ increases monotonically, so that
the radial derivative is positive and roughly constant, i.e. the
signature of solid body rotation. The kinematical parameters
$A(R)$, $B(R)$, $\kappa(R)$, and $\Omega(R)$, as functions of
galactocentric radius, are shown in Figure 4 ({\it bottom}).

\begin{figure}
\epsscale{1.2} \plotone{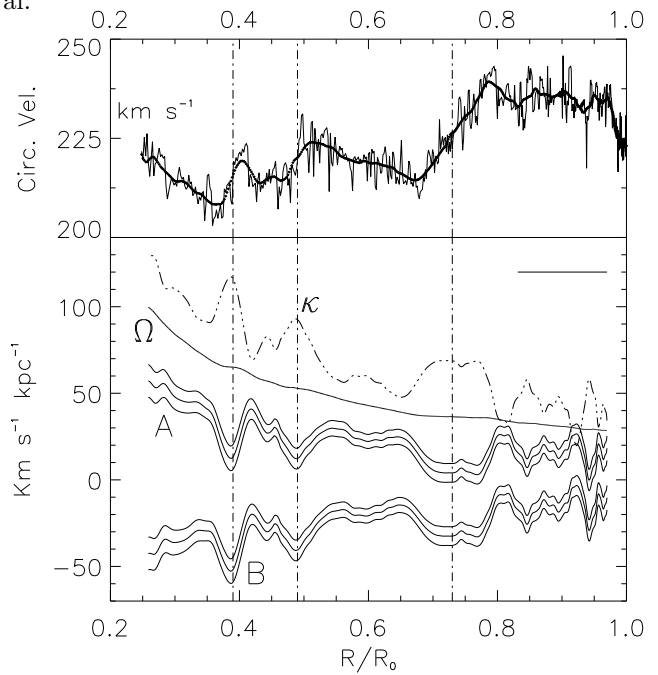} \caption{ Rotation curve and
derived kinematical parameters. The vertical lines depict the
center of the regions presumably with solid-body rotation  {\it
Top}:  A box size of 0.5\,kpc has been used to obtain  a  boxcar
smoothed version (thick line) of the measured rotation curve (thin
line). {\it Bottom}: The epicycle frequency $\kappa(R)$, the
angular velocity $\Omega(R)$, and the Oort parameters $A(R)$ and
$B(R)$, derived from the smoothed version of the rotation curve.
The $1\sigma$ errors in  $A(R)$ and $B(R)$  are shown with thin
lines. The segment  from 0.83 to 0.97 $R/R_0$ (shown at  $120
km\,s^{-1}$ in the plot)  is the radial region used to fit  the
Oort constants A and B. \label{fig4}}
\end{figure}

The angular velocity, $\Omega(R)$, the epicyclic frequency,
$\kappa(R)$, and the parameter $A(R)$, describing the shear,
decrease on the average with galactocentric radius. On the
contrary, the parameter $B(R)$, describing the vorticity, grows
with galactocentric radius. The shear and vorticity present three
relative minima, coincident with local maxima of the epicyclic
frequency $\kappa(R)$, at radii 0.39, 0.47 and 0.73 $R/R_0$. The
parameter $A(R)$, directly proportional to the radial derivative
of the angular velocity, tends to zero at these radii, so the
angular velocity is almost constant, a characteristic of solid
body rotation. The relative maxima in $\kappa(R)$ at these three
radii can be interpretated also as evidence for solid body
rotation; at a given galactocentric radius, $\kappa$ would be
$\sqrt{2}$ times higher for solid body rotation than for a  flat
rotation curve  $V(R) = V_0$ = constant.

The results of our analysis are consistent with other work  in
that the parameters $A(R)$ and $B(R)$  at  $R = R_0$,  are very
close to the values of the Oort's constants  $A_0$ and $B_0$
recommended by the IAU \citep{kerrlynde86}. The values derived
here are obtained from linear fits to $A(R)$ and $B(R)$, within
the range 0.83 to $0.97 R/R_0$, extrapolated to $ R/R_0 = 1 $.
Our results yield $A_0 = 14.9 \pm 4$
km\,s$^{-1}$\,kpc$^{-1}$, and $B_0 = -12.3 \pm 4$
km\,s$^{-1}$\,kpc$^{-1}$, in good agreement with those
recommended by the IAU, $A_0 = 14.5 \pm 2$
km\,s$^{-1}$\,kpc$^{-1}$ and $B_0 = -12.0 \pm 3$
km\,s$^{-1}$\,kpc$^{-1}$. In a similar manner a value of $\kappa_0
= 35 \pm 7$km\,s$^{-1}$\,kpc$^{-1}$ is obtained for the epicyclic
frequency at  $ R = R_0$, in agreement with the value of $\kappa_0
= 36 \pm 10$ km\,s$^{-1}$\,kpc$^{-1}$ given by \cite{byt87}. The
angular velocity obtained for $ R = R_0$, $\Omega_0 = 27.2 \pm 1$
km\,s$^{-1}$\,kpc$^{-1}$,  is comparable with the value of
$\Omega_p = 26 \pm 4$ km\,s$^{-1}$\,kpc$^{-1}$, derived for the
spiral pattern angular velocity (Fern\'andez, Figueras, \& Torra
2001).  Our results therefore suggest that the Sun location is
close to the corotation radius (see also discussion in Amaral \&
Lepine 1997).

\subsection{The subcentral vicinity}

To study the possible relationships between the kinematical
characteristics of the Galaxy, the molecular gas density distribution
and the massive star formation rates (MSFR),  it is necessary to use consistent
data sets.  The densities and MSFRs should be estimated, preferentially,
for the very same regions for which the kinematics have been inferred.
The rotation curve is obtained using information from the subcentral
points of the disk, so the densities and MSFRs we wish to examine
are evaluated in regions adjacent to the subcentral points.  Such
procedure advantageously minimizes the two-fold  ambiguity in
kinematical distance occurring within the solar circle since, at the
subcentral points, kinematical distances are univocally defined \citep{byt87}.

Limiting the analysis to a reduced area of the longitude-velocity
space avoids also the azimuthal averaging of physical conditions
which pertain both to spiral arms and to inter-arm regions, a
difficulty that appears in axisymmetric analysis of the entire
longitude-velocity space. Azimuthal averaging, over disk annuli,
of the gas surface density and star formation rates, allows
derivation of the star formation activity threshold in disk
galaxies \citep{myk01, bossier03}; the detailed information on the
spiral structure of the disk, however, is washed out in the
azimuthal average.

The ``{\it subcentral vicinity}'', used to inspect the gas
physical conditions and MSFR,  is  defined here, at each galactic
longitude, as the velocity span between the terminal velocity and
the terminal velocity plus $\Delta V = 15$\,km\,s$^{-1}$ (Fig. 5).
The value of $15$\,km\,s$^{-1}$ is large enough to include all the
emission from each CO profile component used to select a terminal
velocity. Were the emission to be interpreted as originated by a
set of clouds at different velocities, the range $\Delta R $
implied by $\Delta V = 15$\,km\,s$^{-1}$  would be  0.5 kpc at $l
= 307^\circ$ and 0.2 kpc at $l =342^\circ$, roughly the longitude
limits of the analyzed region. These values of $\Delta R $ are
small compared to the large scale trends with respect to
galactocentric radius and therefore do not affect our conclusions.

\begin{figure}
\epsscale{1.3} \plotone{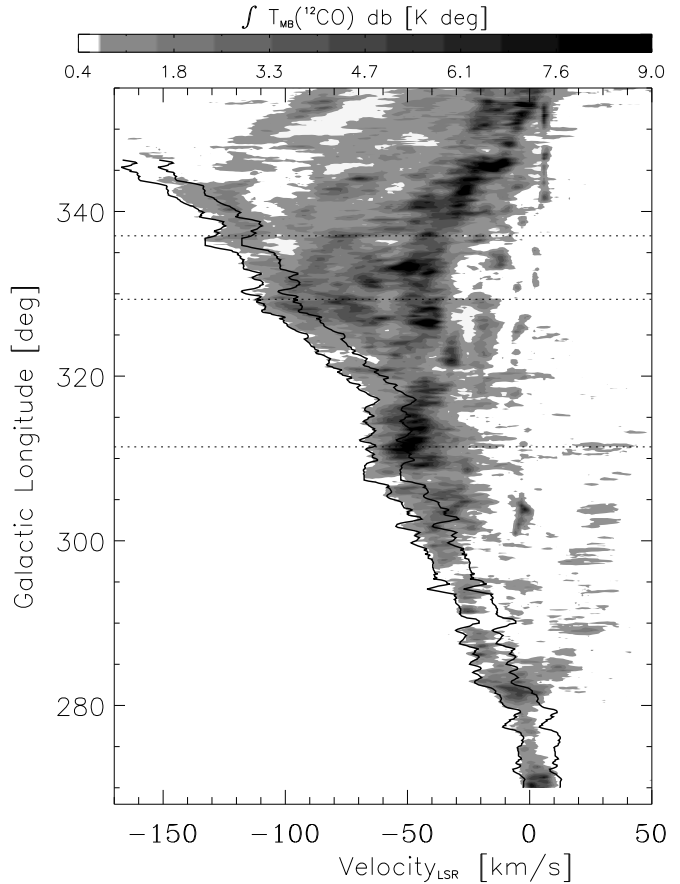} \caption{The subcentral vicinity
borders, separated by
 $\Delta V = 15 km\,s^{-1}$ are shown with the thick lines atop the
longitude-velocity diagram of CO emission (gray scale), integrated
in latitude over $-2^{\circ}\leq b \leq 2^{\circ}$. Horizontal
dotted lines mark the regions with solid body rotation, as in
Figure 4. \label{fig5}}
\end{figure}

The properties of the ISM, molecular gas face-on surface density
and MSFR per unit area, are averaged within the subcentral
vicinity, to evaluate their dependance on galactocentric radius.
It is assumed that the gas shares the kinematics of the
corresponding subcentral point at each longitude.  The molecular
gas face-on surface density is computed from the same CO database
used to derive the rotation curve, while the MSFR is computed from
the FIR luminosity of the IRAS/CS sources.

\subsection{Molecular gas surface density}

The derivation of molecular clouds mass from our CO survey data is
based on the widely made assumption that the velocity-integrated
CO intensity is directly proportional to the total H$_2$ column
density in molecular clouds, and is therefore proportional (with a
correction to account for helium) to the total molecular gas mass
column density. \citep{byt87}.  To evaluate the molecular gas
surface density, $\Sigma_{gas}$, the CO emission is binned in
galactocentric annuli of thickness 0.05\,$R/R_0$ and, for each
radial bin, integrated over all velocities within the subcentral
vicinity and over $[-2^\circ,2^\circ]$ in latitude.  The resultant
quantity is divided by the sampled face-on area. The
proportionality factor $X$ adopted, where N(H$_2) = X W(^{12}$CO),
is equal to $1.56 \times 10^{20}$ cm$^{-2}$
(K\,km\,s$^{-1})^{-1}$, \citep{hunter97}.  A correction of 1.36 is
used to account for helium (see Murphy \& May 1991).  The results
obtained for the gas surface density in the subcentral vicinity,
within the solar circle, are shown in Figure 6 ({\it top}).

\begin{figure}
\epsscale{1.2}
\plotone{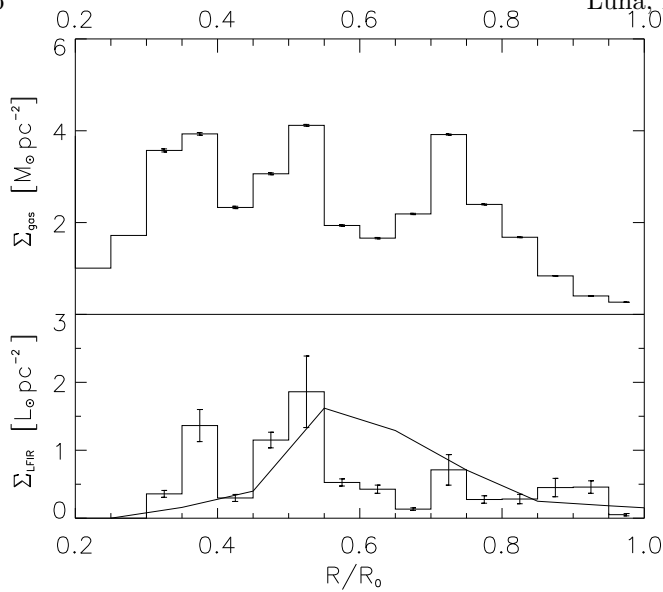} \caption{Molecular gas surface
density and integrated FIR luminosity from massive star forming
regions. {\it Top}: The  face-on surface density  for the
subcentral vicinity. {\it Bottom}: Integrated FIR luminosity per
unit face-on area for the subcentral vicinity compared with an
axisymmetric analysis of the southern galactic disk
\citep{bronf00}. The analysis of the subcentral vicinity shows
structure related with the spiral pattern.
  \label{fig6}}
\end{figure}

\subsection{Rate of massive star formation}

Massive stars form in the dense cores of giant molecular
clouds. The UV radiation from the young stars heats the
surrounding dust, which re-radiates the energy principally in the
far infrared.  Most of these cores are detected as IRAS point-like
sources, with FIR colors typical of ultra-compact H{\small II}
regions \citep{wyc89}. The density is high enough for the
excitation of the CS line, above $10^4$ - $10^5$\,cm$^{-3}$
\citep{bronfnyman}. Following \cite{kennic98a}, the rate of massive
star formation as a function of galactocentric radius is computed by
adding the FIR luminosity of the IRAS/CS sources that fall within
each radial bin.

\begin{equation}
MSFR_{FIR}=6.5 \times
10^{-10}\left(\frac{L_{FIR}}{L_{\odot}}\right) [M_\odot yr^{-1}].
\end{equation}

\noindent The FIR luminosity is evaluated as
$L_{FIR}=4\pi D^2 F_{FIR}$ \citep{Boul88}, where $D$ is the
subcentral distance,

\begin{equation}
F_{FIR} = \sum^4_{i =1} \nu F_\nu(i),
\end{equation}

\noindent  and $F_\nu(i)$ are the FIR fluxes of the IRAS/CS
point-sources in the four IRAS bands, as published in the
Point Source Catalog. The galactic FIR face-on surface luminosity
of IRAS/CS sources as a function of galactocentric radius, for the
subcentral vicinity, is presented in Figure 6 ({\it bottom}), compared
with  the axisymmetric analysis of the complete southern galactic
disk by \citet{bronf00}.

The FIR surface luminosity estimated here for massive star forming
regions constitutes only a lower limit, since the emitting regions
powered by embedded massive stars can be more extended than the
IRAS resolution.  But deconvolution of the extended FIR continuum
emission can be much more difficult because of superposition of
sources along the line of sight. For point-like sources there is
no such confusion since all the CS profiles measured have only one
velocity component \citep{bronfnyman} and, therefore, the
derivation of kinematical distances is straightforward.

The derived MSFR per unit area (Figure
7{\it c}) has relative maxima at the same galactocentric radial
regions than the molecular gas surface density (Fig. 7{\it b}).
These radial regions are characterized by solid-body like
rotation, as shown in the smoothed version of the rotation curve
displayed in (Fig. 7{\it a}).

\subsection{Gravitational disk stability}

The Toomre criterion for disk stability \citep{toomre64, byt87},
used here, is described through the $Q$ stability parameter for
gas, defined by
\begin{equation}
Q(R)=\frac{ \alpha c \kappa(R)}{\pi G \Sigma_{gas}(R)},
\end{equation}
where $c$ is the velocity dispersion, G is the gravitational constant,
$\Sigma_{gas}$ is the gas surface density and $\alpha$ is a
dimensionless parameter that accounts for deviations of real disks
from the idealized Toomre thin disk, single fluid model (Tan
2000).  When $Q < 1$ a gaseous disk is gravitationally unstable.
The particular case $Q = 1$ defines a critical surface density value,
governed by the Coriolis force represented by $\kappa(R)$;

\begin{equation}
\Sigma_{crit}(R)=\frac{\alpha c \kappa(R)}{\pi G}.
\end{equation}

\noindent The Toomre parameter $Q$ can be expressed, hence, as $Q
= \Sigma_{crit}/\Sigma_{gas}$.  The constant $\alpha$ is usually
estimated defining $Q = 1$ where massive star formation ceases
along disks of galaxies \citep{kennic98b, hunter98, Pisano2000}.
Here a value of $\alpha = 0.08$ is found by defining $Q = 1$ in
the galactocentric radial range $R/R_0 = 0.4$ to 0.45, where the
MSFR per unit area is close to zero. Following
\citep{kennic98b}, a value of 6\,kms$^{-1}$ is adopted for the
velocity dispersion. The dependance of $Q$ on galactocentric
radius is shown in Figure 7{\it d}.

To analyze the case of cloud destruction governed by the shear
rate instead of the Coriolis force, Toomre's criterion has been
modified by \citet{elmegreen93}. The shear rate is described by
the Oort parameter $A(R)$; a new parameter, $Q_{A}$,  that
evaluates the survival of a cloud, is in this case

\begin{equation}
Q_{A}(R)=\frac{ \alpha_A 2.5 cA(R) }{\pi G \Sigma_{gas}(R)},
\end{equation}
\noindent Clouds are sheared to destruction, in differentially
rotating disks, for gas surface densities larger than the critical
surface density $\Sigma_{crit}^A$

\begin{equation}
\Sigma_{crit}^A(R)=\frac{2.5\alpha_A c A(R)}{\pi G},
\end{equation}

\noindent so that $Q_{A} =  \Sigma_{crit}^A/\Sigma_{gas}$.  This
modified stability criterion takes into account the important role
of shear in the destruction of GMCs, and can be used to describe
how shear controls the rate of star formation in galactic disks in
the regions where the Coriolis force does not. The parameter
$\alpha_A$ is found, as $\alpha$,  to be  $\alpha_{A} = 0.2$. The
dependance on galactocentric radius of $Q_{A}$  is shown in Figure
7{\it e}.

 A complementary analysis of the stability of the gas clouds to tidal shear
is proposed by \citet{kenney93}, who defines a critical tidal surface density $\Sigma_{tide}$.

\begin{equation}
\Sigma_{tide}(R)=\frac{\sigma_z(R)[3A(A-B)]^{1/2}}{\pi G},
\end{equation}

\noindent where $\sigma_z$ is the velocity dispersion in the
z-direction of the disk.  If the gas density is less than
$\Sigma_{tide}$, tidal shear will rip apart the clouds. The
relative importance of tidal shear and gravitational stability can
be expressed by the ratio of $\Sigma_{tide}$ to the critical
surface density for gravitational instabilities as defined by
Toomre for an ideal disk,

\begin{equation}
\Sigma_{grav}=\frac{ c \kappa}{\pi G}.
\end{equation}
\noindent Assuming that $c~\approx~\sigma_z$, the ratio depends
almost entirely on the shape of the rotation curve,

\begin{equation}
\frac{\Sigma_{tide}}{\Sigma_{grav}}=0.87\left(\frac{-A}{B}\right)^{1/2},
\end{equation}

\noindent and tells whether tidal or gravitational forces determine the disk
stability. The dependance of $\Sigma_{tide}$/$\Sigma_{grav}$ in
galactocentric radius is shown in Figure 7{\it f}.

\subsection {Disk stability, gas density, and massive star formation}

\begin{figure}
\epsscale{1.3} \plotone{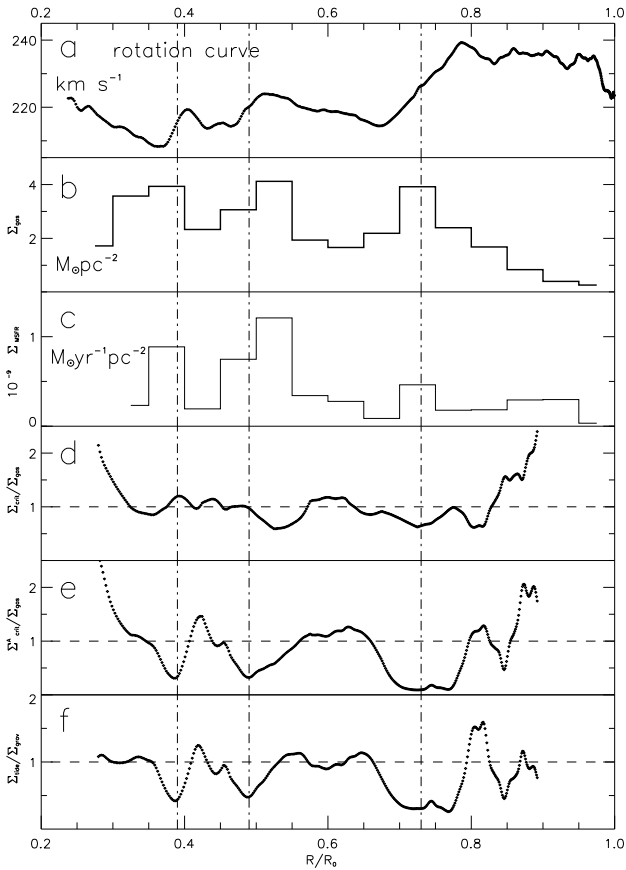}

\caption{Summary of the results from our analysis of the
subcentral vicinity.  The regions with solid-body like kinematics
are shown by vertical dash-dotted lines. Correlations are
discussed in the text. ({\it a}) Smoothed rotation curve. ({\it
b}) Molecular gas face-on surface density ({\it c}) Massive star
formation rate per unit area  $\Sigma_{MSFR}$. ({\it d})
Gravitational instability parameter $\Sigma_{crit}/\Sigma_{gas}$.
({\it e}) shear instability parameter
$\Sigma^A_{crit}/\Sigma_{gas}$. ({\it f}) Parameter
$\Sigma_{tide}/\Sigma_{grav}$ comparing the relative influences of
gravity and tide in the disk stability. \label{fig7}}
\end{figure}

Some of our most relevant results are displayed in Figure 7, which
we now describe in more detail. The values plotted and their
errors are also listed in Table 1.  Panel\,{\it a}, depicts the
smoothed rotation curve used to derive the Oort parameters $A(R)$
and $B(R)$. Uncertainties in the rotation curve, of
$\pm2$\,km\,s$^{-1}$, are dominated by the deviation of the
measured subcentral velocities from the smoothed curve.
Panel\,{\it b} presents the molecular gas face-on surface density
evaluated in galactocentric radial bins of extent $0.05 R/R_0$,
with the first bin centered at $R/R_0 = 0.275$. Panel\,{\it c}
shows the MSFR, in units of solar masses
per year per square pc ($\Sigma_{MSFR}$), averaged in the same
galactocentric radial bins as for {\it b}.  Uncertainties are
estimated from Poisson statistics. Panel\,{\it d} displays the
gravitational stability parameter $Q$ defined by Toomre;
panel\,{\it e} the $Q_A$ parameter defined by \citet{elmegreen93}
for the case when shear dominates over Coriolis force; and
panel\,{\it f}, finally, the ratio of the tidal to the gravitation
surface instabilities (eq. [11]). The ratio {\it f} is estimated
from the parameters $A(R)$ and $B(R)$ derived from observed
rotation curve.

The gas surface density and the MSFR are higher than average, as
apparent from Figure 7, in regions where the rotation is typical
of a solid body, roughly indicated by the vertical lines at radii
0.39, 0.47 and $0.73 R/R_0$ (see also Fig. 4). These three regions
are very close to the accepted values for the tangent points of
spiral arms identified in the literature (e.g. Vall\'ee 2002 and
references therein); the 3\,kpc arm approximately at $0.36 R/R_0$,
the Norma spiral arm at $0.51 R/R_0$ and the Crux spiral arm at
$0.77 R/R_0$. Figure 7{\it d} tells us that although gravitational
instabilities are present in the Galactic disk, the disk is
self-regulated in the sense that $Q$ is of order 1 everywhere
\citep{tan00}. However, as shown in Figure 7{\it e}, where the
Coriolis force is replaced by shear as the principal agent of
cloud destruction, the three regions coincident with spiral arm
tangents are characterized by relative minima. This result is
reinforced by the three relative minima in Figure 7{\it f}, where
the tidal shear is clearly lower than that needed to disrupt the
clouds and, therefore, allows them to survive enough so as to
collapse gravitationally and ultimately form massive stars.

To estimate the influence of the derived quantities and parameters
in the process of massive star formation, we examine the
correlation between $\Sigma_{MSFR}$ (Panel\,{\it c}) and
$\Sigma_{gas}$, $\Sigma_{crit}$/$\Sigma_{gas}$,
$\Sigma_{crit}^A$/$\Sigma_{gas}$, $\Sigma_{tide}/\Sigma_{grav}$,
$\Omega$ and $\kappa$.  A Spearman rank-order correlation
coefficient test \citep{numrecip} is adopted; the ranking and
significance of each correlation are listed in Table 2.  Massive
star formation is clearly correlated with the gas surface density,
with a confidence level larger than $90\%$, and less clearly
correlated with the angular velocity and the epicyclic frequency,
with a confidence level larger than $75\%$ and lower than $85\%$.
For the variables $\Sigma_{crit}^A$/$\Sigma_{gas}$ and
$\Sigma_{tide}$/$\Sigma_{grav}$, which take into account shear,
there is no correlation with respect to the NULL hypothesis that
the dependent variables are drawn from a random distribution.
These results are discussed in the following section, searching
for a possible scenario to explain the observed correlations.

\section{DISCUSSION}

\subsection{Spiral arms in the Southern Galaxy}

There are regions in the rotation curve (Fig. 7{\it a}) that
present solid-body like kinematics; these regions are coincident
with the loci of known spiral arm tangents. The molecular gas
surface density is at least a factor of 2 larger for the arm
regions than for the inter-arm regions (Fig. 7{\it b}).  In those
regions where the rotation is typical of a solid body, the rate of
massive star formation  per unit area, $\Sigma_{MSFR}$,
presents relative maxima (Fig. 7{\it c}). The relative maxima in the
molecular gas surface density and  MSFR are coincident with
relative minima in the kinematical parameters
$\Sigma_{crit}^A$/$\Sigma_{gas}$ and
$\Sigma_{tide}$/$\Sigma_{grav}$.  Such results, as shown in Figure
6{\it b}, cannot  be obtained alone from an axisymmetric analysis
of the Galactic disk.

In those radial regions where the kinematics resembles that of a
solid body, i.e., spiral arm tangents, marked with vertical lines
in Figure 7,  $V/R = \Omega\approx$constant and $-A/B\ll 1$.
Stability analysis suggests that molecular gas  in regions
characterized by solid body rotation is stable
($\Sigma_{gas}\approx\Sigma_{crit}$) or self-regulated
\citep{tan00}; GMCs are not destroyed by tidal disruption
($\Sigma_{tide}\ll\Sigma_{gas}$), and there is an enhancement
in $\Sigma_{MSFR}$. In regions that deviate from solid
body rotation, the star formation decreases, implying that shearing
motions may disrupt/destroy clouds and, therefore, play an important
role in regulating large scale star formation in the disk.

The destruction of clouds by shearing motions, therefore, is
an important parameter controlling the large scale star formation
in the disk, particularly in the inter-arm regions where shear dominates
gravitational stability. High tidal disruption appears to be the
agent for GMC destruction in the inter-arm regions, inhibiting the
formation of massive stars, while low tidal disruption facilitates
the accumulation of gas in the arms, increasing the gas surface
density locally and promoting the higher observed MSFR. Regions of
solid body rotation should present less collisions due to the low
shearing motions and, therefore, the MSFR could presumably be
limited by a process different from shear.

The relation between solid body motion and enhanced star formation
has been also found for external galaxies. A nearby example of a
galaxy in which shear instability may regulate star formation is
M33 \citep{corbelli03, heyer04}. Using a shear stability
criterion, they find that the predicted outer threshold radius for
star formation is consistent with the observed drop in the
H$\alpha$ surface brightness. Another example is the central
region of NGC\,3504. The molecular gas kinematics are derived by
\citet{kenney93}, and compared with the rate of star formation
obtained from the H$\alpha$ recombination line. The central region
of NGC\,3504 is characterized by solid body kinematics and high
rate of star formation; they propose that the behavior of star
formation is strongly influenced by the strength of tidal shear,
which can help control the star formation rate via cloud
destruction. For irregular galaxies, \citet{hunter98} notice that
in slowly raising rotation curves, characteristic of this type of
galaxies, the stability parameter derived by \cite{elmegreen93}
(eq. 8), provides a good criterion to describe the boundary
between cloud survival and disruption. The conclusions of these
authors are similar to our results for the Southern Milky Way
arms, which show solid body kinematics, i.e. that the SF is
strongly influenced by tidal shear, and in particular that tidal
shear controls the rate of star formation through cloud
destruction.

Theoretical work in agreement with the present results has been presented
by \citet{RYS87}. They use a galactic disk model where the ISM is
simulated by a system of particles, representing clouds, which
orbit in a one arm spiral-perturbed gravitational field, and include
dissipative cloud-cloud collision.  Their conclusion is that the
distribution of GMCs and star formation is enhanced across the
full finite width of the spiral arm and is not restricted to either the
preshock or postshock regions.

The agreement between their theoretical results and our
observations implies that the best tracer for molecular gas
spiral structure is the kinematical parameter that compares
gravitational stability and cloud shear destruction
($\Sigma_{tide}/\Sigma_{grav}$, Fig. 7{\it f}), and is also the
simplest one, because it involves only kinematical parameters that
can be derived directly from the rotation curve. Spiral arm
regions can be identified, therefore, as those where shear
disruption of clouds has less influence than gravitational
instabilities, so that molecular clouds  can pile up, increasing
their mass and evolving, until they collapse to form stars that
will thereby increase the mean kinetic temperature. Such increment
in temperature in the spiral arm regions has been preliminarily
measured  by \citet{aluna04}.

\subsection{Massive star formation and the Schmidt Law}

There is a good correlation, as shown in section 3.7, between
the MSFR per unit area (Fig. 7{\it c})
and the molecular gas face-on surface density (Fig. 7{\it b})
in the Milky Way.  Such kind  of correlation, formally known
as Schmidt law, has been explored mostly for external galaxies.
In this section we evaluate, for the subcentral vicinity of the Milky Way
disk,  four different enunciations of the Schmidt Law, whose simplest
form \citep{kennic89} is, in our case,

\begin{equation}
\Sigma_{MSFR} \propto(\Sigma_{gas})^{n},
\end{equation}

Previous efforts at deriving the Schmidt Law for the Milky
Way include those of \citet{guibert78}, who find values of n
between 1.3 and 2.0, based on the distribution of OB associations,
H II regions, H I, and early CO surveys;  and  \citep{bossier03},
who using known gaseous and stellar radial profiles, find that the
Schmidt Law is satisfied for the Milky Way from R = 4 to 15 kpc.
Recently \citep{krumMckee05}, in a general theory of
turbulence-regulated star formation in galaxies, predict the star
formation rate as a function of galactocentric radius for the
Milky Way taking into account the fraction of molecular gas in the
form of clouds, and enunciate Schmidt Law in the assumption that
the clouds are virialized and supersonically turbulent.

A second form of the Schmidt law, which modulates the gas density
by a kinematical parameter, in this case the angular velocity
$\Omega$ \citep{kennic89}, is given by

\begin{equation}
\Sigma_{MSFR}\propto\Sigma_{gas}\Omega.
\end{equation}

A third derivation of the Schmidt Law  \citep{tan00}, can be
expressed as

\begin{equation}
\Sigma_{MSFR}\propto\Sigma_{gas}^{n}\Omega(1-0.7\beta),
\end{equation}

\noindent with

\begin{equation}
\beta \equiv \frac{d [\ln (v_{circ})]}{d [\ln (R)]},
\end{equation}

\noindent where $v_{circ}$ is the circular velocity at a
particular galactocentric radius. This version of Schmidt law
has the advantage that the proportionality constant can be derived
from theory, and includes evaluation of the mean free path for
cloud-cloud collision, the fraction of collisions that lead to
star formation, and the fraction of gas transformed into stars in
each collision.

A fourth derivation, of theoretical character,  that assumes molecular
clouds to be virialized and supersonically turbulent, has been given by
\citet{krumMckee05}, and is expressed here as

\begin{equation}
\Sigma_{MSFR}\propto f_{GMC} \Sigma_{gas}^{0.68}\Omega^{1.32},
\end{equation}

\noindent where $f_{GMC}$ is the  fraction of gas in the form of molecular
clouds

The four enunciations of Schmidt Law are tested (Figure 8) using
the values of molecular gas face-on surface density and of the
MSFR per unit area obtained here for galactocentric radial bins in
the Milky Way. Although most derivations of Schmidt Law include
both atomic and molecular gas, our calculations are limited to the
molecular gas component of the ISM.  \citet{wong02} have found,
for a set of seven galaxies, that the azimuthally averaged star
formation rate per unit area correlates much better with
$\Sigma{_{H_{2}}}$ than with $\Sigma_{HI}$. One should note that
they required a large, uncertain correction for extinction to
derive the rate of star formation using H$\alpha$ images; however,
such correction does not affect their main conclusion, i. e., that
considering the total gas density, rather than H$\small{I}$ and
H$_{2}$ separately, may obscure underlying physical processes that
are essential to star formation. The situation may be presumably
different in galaxies where the ISM is predominantly atomic;
however, even for M33, with a predominantly atomic ISM,
\cite{heyer04} has shown that the star formation rate per unit
area still correlates better with $\Sigma{_{H_{2}}}$ than with
$\Sigma_{HI}$.

  \begin{figure}
\epsscale{1.25} \plotone{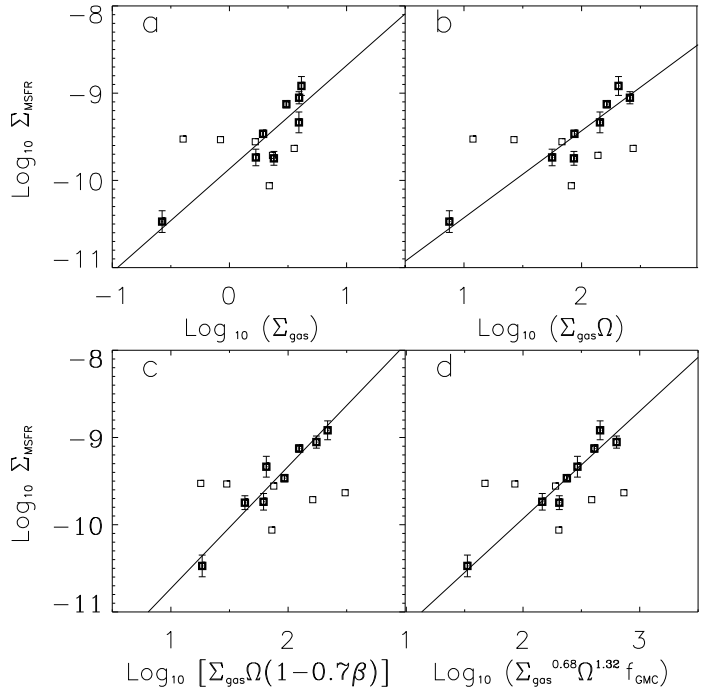}

\caption {The Schmidt law, relating massive star formation rate
and gas density in the subcentral vicinity of the Milky Way,
obtained with four different methods. Thick squares represent
spiral arm tangent regions and thin squares interarm regions.
Linear fits to the radial bins corresponding to spiral arm regions
yield  tight correlations in all four cases (see text). ({\it a})
Simplest expression: $Log(\Sigma_{MSFR} [M_{\odot}yr^{-1}pc^{-2}$)
vs $Log(\Sigma_{gas} [M_{\odot}pc^{-2}$). ({\it b}) Expression
including the angular velocity explicitly, $Log(\Sigma_{gas}
\Omega(R)[M_{\odot}yr^{-1}pc^{-2}kms^{-1}kpc^{-1}]$). ({\it c})
Expression derived by \citet{tan00}. ({\it d}) Expression derived
by \citet{krumMckee05}}

\end{figure}

The Schmidt law is roughly satisfied in all four cases analyzed,
with a power index close to 1.  The dispersion of the data points
is large if all radial bins are taken into account; however, the
relation becomes much tighter if the radial bins that correspond
to interarm regions, shown in Figure 8 as thin squares, are
removed.  For the simplest relation (Fig. 8{\it a}),
$\Sigma_{gas}\propto(\Sigma_{MSFR})^n$, the index value $n=1.2 \pm
0.2$ obtained is in good agreement with those derived for
extragalactic disks \citep{wong02}. The relation plotted at Figure
8{\it b}, taking into account the kinematics through the angular
velocity $\Omega$,  with an index value $n=1.0 \pm 0.1$, is  in
very good agrement with \citet{kennic98b}.  Because they
derive an index from disk-averaged values for a set of galaxies,
rather than the variation of star formation rate with surface density
within a galaxy, presumably describing quite different processes, our
results show an underlaying physical link between $\Sigma_{gas}$
and $\Sigma_{MSFR}$ that dominates both local and global scales in
the process of star formation. The relation for the Schmidt law
at Figure 8{\it c}, dependent on the kinematics, as proposed by
\citet{tan00}, shows a relation with an index value $n=1.4
\pm 0.2$, consistent with theory. An index of  $1.2 \pm 0.1$ is
obtained when fitting the relation derived by \citet{krumMckee05},
using a value of $f_{GMC}=0.25$ \citep{dame87}.

It is apparent that the simplest form (Fig. 8{\it a}) of Schmidt law,
that normally describes the averaged properties of disks of
galaxies \citep{kennic89}, applies also fairly well to $\Sigma_{MSFR}$
in a scale range which is much smaller, like the radial bins in the Milky Way.
The galactic disk kinematics, through shear,
additionally modulates the global star formation in the Galactic
disk; destroys the clouds in the interarm regions and permits the
piling up of gas in the spiral arms, allowing gravity or another
agents like compression and turbulence (via supernovas),
to collapse the clouds into stars (see recent reviews by
\citet{elmegreen02,maclow04}.

\subsection{ Timescale for the growth of gravitational instabilities}

The gas depletion timescale due to star formation, $\tau_{SF}$,
can be crudely estimated from the derived MSFR per unit area,
$\Sigma_{MSFR}$, and the molecular gas surface density,
$\Sigma_{gas}$,

\begin{equation}
\tau_{SF}=\frac{\Sigma_{gas}}{\Sigma_{MSFR}}.
\end{equation}

\noindent The galactocentric trend of $\tau_{SF} $ is shown in
Figure 9 ({\it top}). Its average value is 10$^{10}$\,yr, an upper
limit since $\Sigma_{MSFR}$ represents a lower limit. The
timescale for growth of GMCs through gravitational instabilities
can be expressed as $\tau\sim$\,Q/$\kappa$ \citep{larson88}.
Results here yield timescales of the order $\sim 10^7$yr, with
relative minima at the location of the spiral arm tangents (Fig. 9
{\it middle}). Following \citet{kenney93}, the efficiency of star
formation ($\epsilon$) can be estimated via
$\tau_{SF}$=$\epsilon^{-1} \tau$ if the timescales for
gravitational instabilities ($\tau$) are similar to the timescale
for cloud collapse. The values of a few tenths of a percent
obtained, however, are rather low as compared with values of a few
percent typically reported in the literature.

\begin{figure}
\epsscale{1.2} \plotone{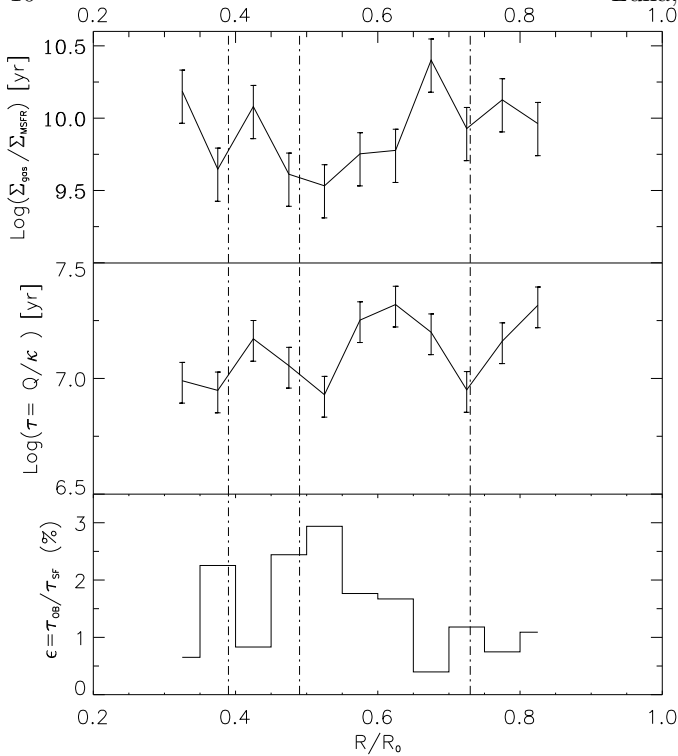}

\caption{Comparison between galactocentric radial variations of
the gas consumption time, the growth of gravitational
instabilities, and the star formation efficiency. Vertical
dash-dotted lines show regions with solid body - like kinematics
(see text). ({\it Top}) Timescale for gas consumption due to star
formation. ({\it Middle}) Timescale for gravitational instability
growth. ({\it Bottom}) Star formation efficiency (percentage)
using a constant 10$^8$\,yr life time  ($\tau_{OB}$) for OB
associations. \label{fig9}}
\end{figure}

The efficiency for massive star formation, according to
\citet{maclow04}, can be further expressed as

\begin{equation}
\epsilon=\tau_{OB}\frac{\Sigma_{MSFR}}{\Sigma_{gas}}=\frac{\tau_{OB}}{\tau_{SF}}.
\end{equation}

\noindent Adopting a typical lifetime for an OB star region of
$\tau_{OB}=10^8$yr and our results for $\Sigma_{MSFR}$ and
$\Sigma_{gas}$ we obtain values for $ \epsilon $ of a few percent
in the spiral arms, and extremely low values for inter-arm
regions, (Fig. 9 {\it bottom}). A comparison of the timescales of
gas consumption and growth of gravitational instabilities, as a
function of galactocentric radius, suggests that fast star
formation is present in regions where rapid cloud growth takes
place.

\section{CONCLUSIONS}

This work analyzes the correlation between molecular gas
kinematical properties, molecular gas surface density,  and rate
of massive star formation in the IV galactic quadrant, using the
most complete data bases available.  The analysis  is carried out
for galactocentric radial bins 0.5 kpc wide, a compromise to avoid
arm-interarm confusion while having good statistics. The data used
are restricted to the subcentral vicinity, to avoid the two-fold
distance ambiguity within the solar circle. The main conclusions
are:

$\bullet$ The rotation curve obtained is similar to that presented
by Alvarez et al. (1990).  Since the sampling in longitude is 4
times denser,  however, it can be used to calculate, as a function
of galactocentric radius, kinematical parameters that require
radial derivatives.

$\bullet$ The angular velocity, $\Omega(R)$, the epicyclic
frequency, $\kappa(R)$, and the parameter $A(R)$ describing the
gas shear, tend to decrease with galactocentric radius; the
parameter $B(R)$, describing the gas vorticity, tends to grow. The
values derived for Oort's constants $A_0$ and $B_0$, at R = $R_0$,
are consistent with those recommended by the IAU.

$\bullet$ Shear and vorticity have relative  minima, and the
epicyclic frequency relative maxima, at radii 0.39, 0.47 and 0.73
$R/R_0$, coincident with the known positions of spiral arm tangent
regions.  Near these radii the kinematics are characteristic of
solid body rotation: $A(R)$, proportional to the radial derivative
of the angular velocity, tends to zero, so the angular velocity is
roughly constant. The relative maxima in epicyclic frequency are
consistent with such scenario since $\kappa$ is $\sqrt{2}$ times
higher for solid body rotation than for a flat rotation curve.

$\bullet$ Differential rotation and shear are weaker for the
spiral arm regions than for the interarm regions. The relative
importance of tidal shear w/r to gravitation in the stability of
the gas for spiral arm regions, where the rotation curve resembles
that of a solid body, is half than for spiral arms, where the
rotation curve is nearly flat.

$\bullet$ Massive star formation occurs in regions of high
molecular gas density, roughly coincident with the three lines of
sight tangent to spiral arms.  In these arms the formation of
massive stars follows the Schmidt law, $\Sigma_{MSFR} \propto
[\Sigma_{gas}]^n$, with an index of $n = 1.2 \pm 0.2$. While this
law is characteristic of spiral galactic disks, here it applies to
much smaller spatial scale.  A modified version of Schmidt law,
which modulates the gas density by the angular velocity,
$\Sigma_{MSFR} \propto\Sigma_{gas}\, \Omega$, describes better the
behavior of the gas at this scale, suggesting that the kinematics,
through shear, regulate global star formation in the Galactic
disk.

\acknowledgments

A.L. and L.C. acknowledge support by CONACYT M\'exico through the
research grants 89964, 42609 and G28586. L.B., J.M., and A.L.
acknowledge support from FONDECYT Grant 1010431 and from the
Chilean {\sl Center for Astrophysics} FONDAP No. 15010003. We
thank Drs. E. Brinks, A. Garcia-Barreto, N. Vera and W. Wall for
fruitful comments. A.L. thanks INAOE's Astronomy Department and is
also grateful for the use of facilities and the support at the
Departmento de Astronom\'{\i}a, Universidad de Chile.
We are also grateful to the referee for several helpful comments.

\clearpage

\begin{table}

\caption{Values of our derived and measured parameters in the
adopted bins at the subcentral vicinity. Our analysis beyond
$0.875 R/R_0$ is uncertain and therefore excluded.}

\begin{tabular}{|c|c|c|c|c|c|c|}
\hline \hline

Gal. radii \tablenotemark{a} & circ. vel. \tablenotemark{b} &
$\Sigma_{gas}$ \tablenotemark{c} & $\Sigma_{MSFR}$(10$^{-9}$)
\tablenotemark{d} & $\Sigma_{crit}/\Sigma_{gas}$ \tablenotemark{e}
& $\Sigma^{A}_{crit}/\Sigma_{gas}$ \tablenotemark{f} &
$\Sigma_{tide}/\Sigma_{grav}$ \tablenotemark{g} \\

\hline

R/R$_0$  & kms$^{-1}$ & M$_{\odot}$pc$^{-2}$  &
M$_{\odot}$yr$^{-1}$pc$^{-2}$  &
& & \\

 & $\pm$2 & & & & & \\

\hline
 0.325 & 212.4 &   3.57 $\pm$0.03 & 0.23 $\pm$0.03 &   0.99 $\pm$0.11 &   1.12 $\pm$0.10  &   1.03 $\pm$0.12 \\
 0.375 & 209.6 &   3.93 $\pm$0.03 & 0.89 $\pm$0.15 &   1.00 $\pm$0.04 &   0.43 $\pm$0.10  &   0.56 $\pm$0.11 \\
 0.425 & 214.2 &   2.33 $\pm$0.03 & 0.19 $\pm$0.03 &   1.07 $\pm$0.14 &   1.45 $\pm$0.11  &   1.17 $\pm$0.18 \\
 0.475 & 215.7 &   3.06 $\pm$0.02 & 0.75 $\pm$0.07 &   1.01 $\pm$0.05 &   0.51 $\pm$0.10  &   0.61 $\pm$0.11 \\
 0.525 & 223.6 &   4.12 $\pm$0.02 & 1.21 $\pm$0.34 &   0.59 $\pm$0.06 &   0.58 $\pm$0.06  &   0.93 $\pm$0.11 \\
 0.575 & 219.3 &   1.94 $\pm$0.02 & 0.34 $\pm$0.03 &   1.09 $\pm$0.11 &   1.13 $\pm$0.11  &   0.97 $\pm$0.13 \\
 0.625 & 218.0 &   1.66 $\pm$0.02 & 0.28 $\pm$0.04 &   1.14 $\pm$0.12 &   1.25 $\pm$0.11  &   1.00 $\pm$0.15 \\
 0.675 & 214.5 &   2.19 $\pm$0.01 & 0.09 $\pm$0.01 &   0.91 $\pm$0.06 &   0.65 $\pm$0.09  &   0.75 $\pm$0.13 \\
 0.725 & 225.3 &   3.92 $\pm$0.01 & 0.46 $\pm$0.15 &   0.63 $\pm$0.01 &   0.09 $\pm$0.06  &   0.31 $\pm$0.21 \\
 0.775 & 236.7 &   2.40 $\pm$0.01 & 0.18 $\pm$0.04 &   0.99 $\pm$0.02 &   0.17 $\pm$0.10  &   0.34 $\pm$0.20 \\
 0.825 & 234.3 &   1.68 $\pm$0.01 & 0.18 $\pm$0.04 &   0.91 $\pm$0.10 &   1.03 $\pm$0.09  &   1.03 $\pm$0.22 \\
 0.875 & 236.1 &   0.84 $\pm$0.01 & 0.29 $\pm$0.09 &   1.69 $\pm$0.20 &   2.03 $\pm$0.17  &   1.08 $\pm$0.26 \\
 0.925 & 233.1 &   0.41 $\pm$0.01 & 0.30 $\pm$0.06 & --   & -- & -- \\
 0.975 & 233.6 &   0.33 $\pm$0.01 & 0.03 $\pm$0.01 & --   & -- & -- \\

\hline

\end{tabular}

\tablenotetext{a}{Central bin position}

\tablenotetext{b}{Circular velocity from the smoothed rotation
curve at central bin position}

\tablenotetext{c}{Averaged molecular gas surface density by bin}

\tablenotetext{d}{Averaged rate of massive star formation, per
unit area,
 by bin}

\tablenotetext{e,f,g}{Stability parameters evaluated using the
interpolated version of averaged surface quantities by bin, and
the related kinematic parameter from the smoothed rotation curve
at each position}

\end{table}

\begin{table}
\caption{The Spearman test to explore possible correlations
between $\Sigma_{MSFR}$ and  $\Sigma_{gas}$, as well as between
$\Sigma_{MSFR}$ and kinematical and instability parameters
presented in Table 1.}

\centering
\begin{tabular}{|c|c|c|c|c|c|c|}
\hline

correlation with $\Sigma_{MSFR}$  &  $\Sigma_{gas}$  &
$\Sigma_{crit}/\Sigma_{gas}$  & $\Sigma^{A}_{crit}/\Sigma_{gas}$ &
$\Sigma_{tide}/\Sigma_{grav}$ &  $\Omega$ &  $\kappa$  \\

\hline

Rank               &   0.53     &  -0.02  &  -0.24  &  -0.26 &  0.41  & 0.36 \\
\hline
confidence level   &  $>$90$\%$  & 7$\%$   &  55$\%$  &  60$\%$ & 82\%  & 76\% \\

\hline
\end{tabular}
\end{table}


\begin{thebibliography}{}
\bibitem[Aalto et al.(1999)]{aalto99} Aalto, S., Huttemeister, S., Scoville, N.\,Z., \& Thaddeus, P. 1999, \apj, 522, 165
\bibitem[Alvarez et al.(1990)]{alvarez90} Alvarez, H., May, J., \& Bronfman, L. 1990, \apj, 348, 495
\bibitem[Amaral et al.(1997)]{Amaral97} Amaral, L. H., \& L\'epine, J. R. D. 1997, \mnras, 286, 885
\bibitem[Binney \& Tremaine(1987)]{byt87}  Binney, J., \& Tremainen S. 1987, Galactic Dynamics (Princeton: Princeton Univ. Press)
\bibitem[Bosssier et al.(2003)]{bossier03} Bossier, S., Prantzos, N., Boselli, A., and Gavazzi, G. 2003, \mnras, 346, 1215
\bibitem[Boulanger \& Perault(1988)]{Boul88}  Boulanger, F., \& Perault, M. 1988, \apj, 330, 964
\bibitem[Bronfman et al.(1988)]{bronf88} Bronfman, L., Cohen, R., Alvarez, H., May, J. \& Thaddeus, P. 1988, \apj, 324, 248
\bibitem[Bronfman et al.(1989)]{bronf89} Bronfman, L., Alvarez, H., Cohen, R., \& Thaddeus, P. 1989, \apjs, 71, 481.
\bibitem[Bronfman et al.(1996)]{bronfnyman}  Bronfman, L., Nyman, L.A., \& May, J. 1996, \aap, 115, 81
\bibitem[Bronfman et al.(2000)]{bronf00}  Bronfman, L., Casassus, S., May, J., \& Nyman, L.A. 2000, \aap, 358, 521
\bibitem[Burton (1988)]{burton88}  Burton, B. 1988, in Galactic and Extragalactic Radioastronomy, ed. G. L. Vershuur \& K. I. Kellerman, (New York: Springer-Verlag) 295
\bibitem[Carrasco \& Serrano(1983)]{cys83} Carrasco, L., \& Serrano, A. 1983, IAU Symp. 100, Internal Kinematics and Dynamics of Ga\-la\-xies, (Dordrech: Reidel), 135
\bibitem[Caswell \& Haynes(1987)]{cyh87} Caswell, J., $\&$ Haynes, R. 1987, \aap, 171, 261
\bibitem[Corbelli(2003)]{corbelli03} Corbelli, E. 2003, \mnras, 342, 199
\bibitem[Dame et al.(1986)]{dame86} Dame, T., Elmegreen, B., Cohen, R., Thaddeus, R. 1986, \apj, 305, 892
\bibitem[Dame et al.(1987)]{dame87} Dame, T., et al. 1987, \apj, 322, 706
\bibitem[Dame et al.(2001)]{dame01} Dame, T., Hartmann, D., \& Thaddeus, R. 2001, \apj, 547, 792
\bibitem[Elmegreen(1993)]{elmegreen93}  Elmegreen, B. 1993, in Star Formation, Galaxies and the Interstellar Medium, ed. Franco, Ferrini, \& Tenorio-Tagle, (Cambridge: Cambridge Univ. Press)
\bibitem[Elmegreen(2002)]{elmegreen02}  Elmegreen, B. 2002, \apj, 577, 206
\bibitem[Evans(1999)]{evans99} Evans, N.J. 1999, \araa 37, 311
\bibitem[Fern\'andez et al.(2001)]{fernan01} Fern\'andez, D., Figueras, \& F., Torra, J. 2001, \aap, 327, 833
\bibitem[Georgelin \& Georgelin(1976)]{gg76} Georgelin, Y.M., \& Georgelin, Y.P. 1976, \aap, 49, 57
\bibitem[Grabelsky et al.(1987)]{grabelsky87} Grabelsky, D.A., Cohen, R.S., Bronfman, L., Thaddeus, P., \& May, J. 1987, \apj, 315, 122
\bibitem[Guibert et al.(1978)]{guibert78} Guibert, J., Lequeux, J., \& Viallefond, F. 1978, \aap, 68, 1
\bibitem[Heyer et al.(2004)]{heyer04} Heyer, M., Corbelli, E., Schneider, S., \& Young, J. 2004 \apj, 602, 723
\bibitem[Hunter et al.(1997)]{hunter97} Hunter, S.D., Bertsch J.R., Catelli J.R. et al. 1997, \apj 481, 205
\bibitem[Hunter et al.(1998)]{hunter98} Hunter, D.A., Elmegreen B.G., Baker A.L. 1998, \apj 493, 595
\bibitem[Kenney et al.(1993)]{kenney93}  Kenney, J., Carlstrom, J. and Young, J. 1993, \apj, 418, 687
\bibitem[Kennicutt(1989)]{kennic89} Kennicutt, R. 1989, \apj, 344, 685
\bibitem[Kennicutt(1998a)]{kennic98a} Kennicutt, R. 1998a, \araa, 36, 189
\bibitem[Kennicutt(1998b)]{kennic98b} Kennicutt, R. 1998b, \apj, 498, 541
\bibitem[Kerr \& Lynden-Bell(1986)]{kerrlynde86}  Kerr, F.J., \& Lynden-Bell D. 1986, \mnras, 221, 1023
\bibitem[Krumholz \& McKee(2005)]{krumMckee05} Krumholz, M.R., \& McKee, C.F. 2005, \apj 630, 250
\bibitem[Larson(1988)]{larson88} Larson, R. B. 1988, in Galactic and Extragalactic star Formation, ed. R. E. Pudritz \& M. Fich (Dordrecht: Kluwer), 435
\bibitem[Luna et al.(2001)]{aluna01} Luna, A., Carrasco, L., Wall, W., \& Bronfman, L. 2001, ASP Conf. Ser. 275, Disks of Galaxies: Kinematics, Dynamics and Perturbations, ed. Athanassoula, Bosma, \& M\'ujica (San Francisco: ASP), 131
\bibitem[Luna et al.(2004)]{aluna04} Luna, A., Wall, W., Carrasco, L., Bronfman, L., \& Hasegawa, T. 2004, II international workshop on science with the GTC: Science with GTC 1st-light instruments and the LMT, held in Cd. M\'exico, Mex., 16-18 febrero 2004, in press
\bibitem[Martin \& Kennicutt(2001)]{myk01} Martin, C.L., \& Kennicutt, R.C. 2001, \apj, 555, 301
\bibitem[Mac Low \& Klessen(2004)]{maclow04} Mac Low, M.M., \& Klessen, R.S. 2004, Reviews of modern physics, 76, 125
\bibitem[Murphy \& May(1991)]{murphymay91} Murphy, D.C., \& May, J. 1991, \aap, 555, 301
\bibitem[Pisano et al.(2000)]{Pisano2000} Pisano, D.J., Wilcots, E. M., \& Elmegreen, B. 2000, \apj, 120, 763
\bibitem[Press et al.(1992)]{numrecip} Press, W.H., Teukolsky, S.A., Veterling, W.T., Flannery, B.R. 1992, Numerical Recipes, The Art of Scientific Computing (Second Edition),  Cambridge University Press
\bibitem[Roberts \& Stewart(1987)]{RYS87} Roberts, W. W., \& Stewart, G. 1987, \apj, 314, 10
\bibitem[Robinson et al.(1983)]{rob83} Robinson, B., McCuthcheon, W., Manchester, R., \& Whiteoak, J. 1983, Surveys of the Southern Galaxy, ed. Burton \& Israel, 1
\bibitem[Sawada et al.(2001)]{sawada01} Sawada, T., et al. 2001, \apjs, 136, 189
\bibitem[Schmidt(1959)]{schm59} Schmidt, M. 1959 \apj, 129, 243
\bibitem[Solomon et al.(1986)]{solomon86} Solomon, P.M., Sanders, D.B, \& Rivolo, A.R. 1986, \apj, 292, L19
\bibitem[Sellwood \& Balbus(1999)]{sellwood99} Sellwood, J. A., \& Balbus, S. A. 1978, \apj, 511, 660
\bibitem[Sinha(1978)]{sinha78} Sinha, R. P. 1978, \aap, 69, 227
\bibitem[Sofue \& Rubin(2001)]{sofuerubin01} Sofue, Y., \& Rubin, V. 2001, \araa, 39, 137
\bibitem[Tan(2000)]{tan00} Tan, J. 2000, \apj, 536, 173
\bibitem[Toomre(1964)]{toomre64} Toomre, A. 1964, \apj, 139, 1217
\bibitem[Vall\'ee(2002)]{valle02} Vall\'ee, J. 2002, \apj, 566, 261
\bibitem[Wong \& Blitz(2002)]{wong02} Wong, T., \& Blitz, L. 2002, \apj, 569, 157
\bibitem[Wood \& Churchwell(1989)]{wyc89} Wood, D.O.S., $\&$ Churchwell, E. 1989, \apj, 340, 265

\end{thebibliography}
\end{document}